\documentclass[groupedaddress,aps,pra,superscriptaddress,showpacs,twocolumn,prl]{revtex4}%
\usepackage{epsfig,dsfont,amssymb,amsmath,amsthm,amsfonts,amsbsy,mathrsfs}
\usepackage{graphicx, color}
\usepackage{epstopdf}
\usepackage{mathdots}
\usepackage{float}
\usepackage{xcolor}
\begin{document}

\title{A new parameterized monogamy relation between entanglement and equality}

\author{Zhi-Xiang Jin}
\affiliation{School of Physics, University of Chinese Academy of Sciences, Yuquan Road 19A, Beijing 100049, China}
\affiliation{Max-Planck-Institute for Mathematics in the Sciences, Leipzig 04103, Germany}
\author{Shao-Ming Fei}
\thanks{Corresponding author: feishm@cnu.edu.cn}
\affiliation{Max-Planck-Institute for Mathematics in the Sciences, Leipzig 04103, Germany}
\affiliation{School of Mathematical Sciences, Capital Normal University,  Beijing 100048,  China}
\author{Xianqing Li-Jost}
\thanks{Corresponding author: xianqing.li-jost@mis.mpg.de}
\affiliation{Max-Planck-Institute for Mathematics in the Sciences, Leipzig 04103, Germany}
\author{Cong-Feng Qiao}
\thanks{Corresponding author: qiaocf@ucas.ac.cn}
\affiliation{School of Physics, University of Chinese Academy of Sciences, Yuquan Road 19A, Beijing 100049, China}
\affiliation{CAS Center for Excellence in Particle Physics, Beijing 100049, China\\ \vspace{7pt}}

\begin{abstract}
We provide a generalized definition of the monogamy relation for entanglement measures. A monogamy equality rather than the usual inequality is presented based on the monogamy weight, from which we give monogamy relations satisfied by the $\alpha$th $(\alpha>0)$ power of the entanglement measures. Taking concurrence as an example, we further demonstrate the significance and advantages of these relations. In addition, we show that monogamy relations can be recovered by considering multiple copies of states for every non-additive entanglement measure that violates the inequalities. We also demonstrate that the such relations for tripartite states can be generalized to multipartite systems.

\end{abstract}
\maketitle

Entanglement is essential in quantum communication and quantum information processing \cite{MAN,RPMK,FMA,KSS,HPB,HPBO,JIV,CYSG}.
Among the properties of entanglement, the monogamy of entanglement is a non-intuitive phenomenon of quantum physics, which characterizes the fundamental differences between quantum and classical correlations and measures the shareability of entanglement in a composite quantum system. From the perspective of monogamy of quantum entanglement, a quantum system entangled with one of the other subsystems limits its entanglement with the remaining subsystems. The study of entanglement and its distribution is of great significance in many areas, such as quantum cryptography \cite{jmr,ml}, phase detection \cite{max,kjm,gsm}, and quantum key distribution \cite{vss,MP}.

Over the last two decades, the precise quantitative formalization of the monogamy of entanglement has attracted much attention.
Given any tripartite state $\rho_{ABC}$ and bipartite entanglement measure $E$, the monogamy relation provided by Coffman, Kundu, and Wootters (CKW) \cite{ckw} is quantitatively displayed as an inequality of the following form:
\begin{eqnarray}\label{m1}
E(\rho_{A|BC})\geq E(\rho_{AB})+E(\rho_{AC}),
\end{eqnarray}
where $\rho_{AB}=\mathrm{Tr}_C(\rho_{ABC})$, and $\rho_{AC}=\mathrm{Tr}_B(\rho_{ABC})$. The inequality (\ref{m1}) states that the sum of the individual pairwise entanglements between $A$ and each of the other parties $B$ or $C$ cannot exceed the entanglement between $A$ and joint system $BC$. Such inequalities are not always satisfied by all entanglement measures for any state. The first monogamy relation was proven for three arbitrary qubit states for a squared concurrence \cite{ckw}. Variations in the monogamy relation and generalizations to arbitrary multipartite systems have been established for a number of entanglement measures, like continuous-variable entanglement \cite{jin,ggy,agf,hai,agi,agd,byk,gyzl}, squashed entanglement \cite{MK,cwa,yd}, entanglement negativity \cite{ofh,kds,hvg,ckj,lly,jzx,ZXN,jzx1}, Tsallis-q entanglement \cite{kjs,kjsg,jll}, and Renyi-entanglement \cite{ksb,cdm,wmv}.

Here, inequality (\ref{m1}) captures only the partial property of this measure of entanglement, as other measures of entanglement may not satisfy this relation. This implies that the inequality is not universal but depends on the specific choice of the measure $E$. In this study, we present a general approach for treating the monogamy of entanglement. This property is given by a family of monogamy relations:
\begin{eqnarray}\label{m2}
E(\rho_{A|BC})\geq f(E(\rho_{AB}), E(\rho_{AC})),
\end{eqnarray}
satisfied for any state $\rho_{ABC}$, where $f$ is a continuous function of variables $E(\rho_{AB})$ and $E(\rho_{AC})$ as mentioned in \cite{lan}.
For convenience, we denote $E(\rho_{A|BC})=E_{A|BC}$ and $E(\rho_{AB})=E_{AB}$.
For a particular choice of the function $f(x,y)=x+y$, we recover the CKW inequality (\ref{m1}) from (\ref{m2}).
If $f(x,y)=\sqrt{x^2+y^2}$, we obtain a monogamy relation satisfied by the squared concurrence. Therefore, one can obtain a family of monogamy inequalities based on the proper choice of the function $f$. Recalling that $E$ is an entanglement monotone, we have $E_{A|BC}\geq \max\{E_{AB}, E_{AC}\}$ since it is non-increasing under partial traces.
Therefore, the entanglement distribution is confined to a region smaller than a square with a side length $E_{A|BC}$.  We say that it is monogamous if a nontrivial function $f$ exists such that the generalized monogamy equality $E_{A|BC}= f(E_{AB}, E_{AC})$ is satisfied for any state $\rho_{ABC}$. To consider all cases of entanglement distribution, we use a linear function to traverse all points in a square with side length $E_{A|BC}$ for simplicity (see Fig. 1).

We consider the function $f(x,y)=x+y$ as a rubber band. For two fixed endpoints $(E_{A|BC},0)$ and $(0, E_{A|BC})$, one can obtain different types of functions by moving the point $(\frac{E_{A|BC}}{2}, \frac{E_{A|BC}}{2})$ to the point $(E_{A|BC}, E_{A|BC})$ or to the origin $(0,0)$, as shown in Fig. 1. For any $0< k\leq E_{A|BC}$ with $(k,k)$, that is, the points on the dotted line in Fig. 1, we have the following trade-off between the values of $E_{AB}$ and $E_{AC}$:
\begin{equation}\label{m3}
    \begin{cases}
    \frac{E_{A|BC}-k}{k}E_{AB}+E_{AC} = E_{A|BC}, ~~ E_{AB}\leq E_{AC}\\[2mm]
       E_{AB}+ \frac{E_{A|BC}-k}{k}E_{AC} = E_{A|BC},~~E_{AC}\leq E_{AB}.
     \end{cases}
\end{equation}
In fact, we mainly consider the range of $\frac{E_{A|BC}}{2}<k\leq E_{A|BC}$ because the traditional monogamy inequality is always satisfied for $0< k\leq \frac{E_{A|BC}}{2}$ (blue region in Fig. 1).
From (\ref{m3}), we can define a monogamy relation without using an inequality as follows.

{\bf Definition}
Let $E$ be a measure of entanglement. $E$ is said to be monogamous if for any state $\rho_{ABC}$, it satisfies 
\begin{eqnarray}\label{m4}
E_{A|BC}= \mu E_{AB}+E_{AC}
\end{eqnarray}
for some $\mu (\mu> 0)$, where $E_{AB}\leq E_{AC}$. We call $\mu$ the monogamy weight with respect to the entanglement measure $E$.

{\it Remark} In Ref. \cite{lan} the authors addressed the generalized monogamy inequality $E(\rho_{A|BC})\geq f(E(\rho_{AB}), E(\rho_{AC}))$, which is satisfied for any state $\rho_{ABC}$ with a nontrivial function $f$. The monogamy relation (\ref{m4}) includes this generalized monogamy inequality as a special case. In general, (\ref{m4}) should be given in the form $E(\rho_{A|BC})= f(\mu E(\rho_{AB}), E(\rho_{AC}))$ with a monogamy weight $\mu$ for any state $\rho_{ABC}$.
For any given function $f$, either $E(\rho_{A|BC})\geq f(E(\rho_{AB}), E(\rho_{AC}))$ or $E(\rho_{A|BC})< f(E(\rho_{AB}), E(\rho_{AC}))$, for any state $\rho_{ABC}$, it is always possible to find some $\mu$ such that $E(\rho_{A|BC})= f(\mu E(\rho_{AB}), E(\rho_{AC}))$. Therefore, the monogamy equality (\ref{m4}) is more powerful than the generalized monogamy inequality. In fact, one can use any function $f$ that traverses all points in a square with side length $E_{A|BC}$, as shown in Fig. 1. For simplicity, we use a linear function that gives rise to (\ref{m4}).

(\ref{m4}) yields a generalized monogamy relation without inequality. The monogamy weight $\mu$ defined in Eq. (4) establishes the connections among $E_{A|BC}$, $E_{AB}$, and $E_{AC}$ for a tripartite state. If $\mu\geq1$, then the monogamy inequality (\ref{m1}) is obviously true from (\ref{m4}). The corresponding entanglement distribution is confined to the blue region, as shown in Fig. 1. The case of $0<\mu< 1$ is beyond the CKW inequality. The corresponding regions of the entanglement distribution are the orange, yellow, and white regions in Fig. 1. Specifically, it reduces to the CKW inequality (\ref{m1}) when $\mu=1$.
When $\mu=0$, we have $E_{A|BC}= E_{AC}\geq E_{AB}>0$, according to definition (\ref{m4}). In this situation, we say that the entanglement measure $E$ is non-monogamous. The corresponding entanglement distribution is located at the boundary of the square, except for the coordinate axis, as shown in Fig. 1.
That is, when $\mu\to 0$, $E$ is not likely to be monogamous. On the contrary, $\mu\to 1$ implies that $E$ is more likely to be monogamous.

\begin{figure}
\centering
\includegraphics[width=8cm]{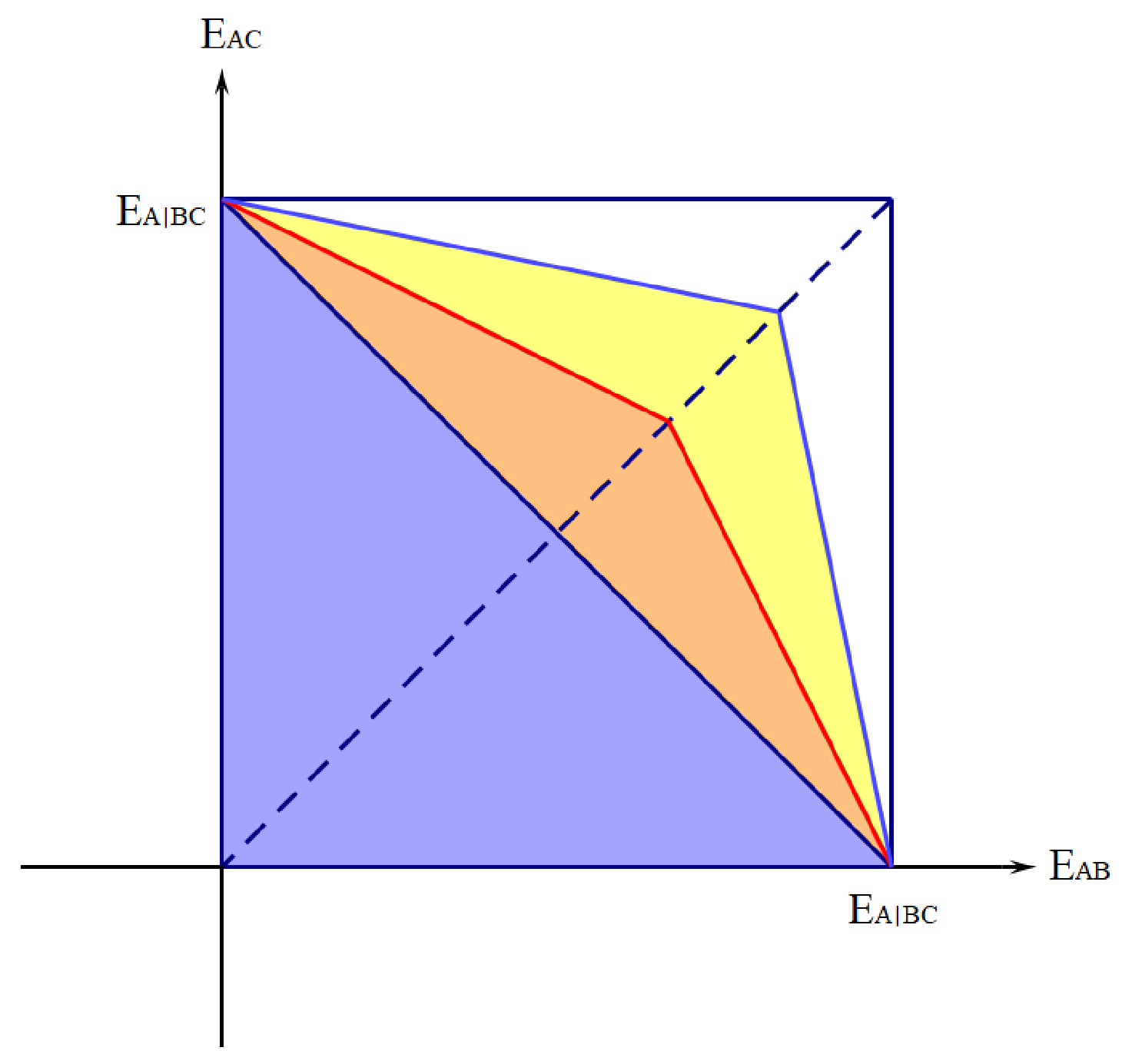}\\
\caption{For any tripartite state $\rho_{ABC}$ and entanglement measure $E$, one gets the CKW inequality (\ref{m1}) for $f(x,y)=x+y$, which holds with the range of values of $E_{AB}$ and $E_{AC}$ given by the blue triangular. In the blue region, the equality (\ref{m4}) also holds for $\mu\geq 1$. In the red, yellow, and white regions, the CKW inequality is no longer satisfied. However, the relation (\ref{m4}) holds for $0<\mu<1$: the orange region matches $\frac{1}{2}\leq \mu <1$, the yellow region matches $\frac{1}{3}\leq \mu < \frac{1}{2}$, and the white region matches $0< \mu <\frac{1}{3}$. In other words, any measure $E$ is monogamous in the sense of (\ref{m4}) if the entanglement distribution is confined to a region strictly smaller than the square with side length $E_{A|BC}$.}\label{2}
\end{figure}

Therefore, the parameter $\mu$ has an operational interpretation of the ability to be monogamous for entanglement measure $E$.
Given two entanglement measures, $E'$ and $\tilde{E}$, with monogamy weights $\mu_1$ and $\mu_2$, respectively, we say that $E'$ has a higher monogamy score than $\tilde{E}$ if $\mu_1\geq \mu_2$ and $E'\succeq \tilde{E}$ or $\tilde{E}\preceq E'$.
Thus, $\mu$ is closely related to the monogamy inequality for a given measure, $E$.
For $E^\alpha$ ($\alpha>0$), we have the following familiar relation (see the proof in the Appendix).

{\bf Theorem 1}  Let $E$ be a measure of entanglement. $E$ is monogamous according to definition (\ref{m4}) if and only if $ \alpha > 0 $ exists, such that
\begin{eqnarray}\label{theorem1}
E^\alpha_{A|BC}\geq E^\alpha_{AB}+E^\alpha_{AC},
\end{eqnarray}
for any state $\rho_{ABC}$.

In the following, we consider the entanglement measure tangle as an application to illustrate the advantages of (\ref{m4}) and the calculation of the monogamy weight $\mu$. The tangle of the bipartite pure state $|\psi\rangle_{AB}\in\mathds{H}_A\otimes \mathds{H}_B$ is given by $\tau(|\psi\rangle_{AB})={2\left[1-\mathrm{Tr}(\rho_A^2)\right]}$ \cite{ckw}.
where $\rho_A=\mathrm{Tr}_B(|\psi\rangle_{AB}\langle\psi|)$ is the reduced density matrix obtained by tracing the subsystem $B$. The tangle for a bipartite mixed state $\rho_{AB}$ is defined by the convex roof extension:
$\tau(\rho_{AB})=\min_{\{p_i,|\psi_i\rangle\}}\sum_ip_i\tau(|\psi_i\rangle)$,
where the minimum is taken over all possible pure-state decompositions of $\rho_{AB}=\sum\limits_{i}p_i|\psi_i\rangle\langle\psi_i|$, where $p_i\geq0$, $\sum\limits_{i}p_i=1$, and $|\psi_i\rangle\in \mathds{H}_A\otimes \mathds{H}_B$.

Let us consider a three-qubit state $|\psi\rangle$ in the generalized Schmidt decomposition form
\begin{eqnarray}\label{ex}
|\psi\rangle&=&\lambda_0|000\rangle+\lambda_1e^{i{\varphi}}|100\rangle+\lambda_2|101\rangle \nonumber\\
&&+\lambda_3|110\rangle+\lambda_4|111\rangle,
\end{eqnarray}
where $\lambda_i\geq0$, $i=0,1,2,3,4$, in descending order and $\sum\limits_{i=0}\limits^4\lambda_i^2=1.$
From this definition, we obtain $\tau_{A|BC}=4\lambda^2_0({\lambda_2^2+\lambda_3^2+\lambda_4^2})$, $\tau_{AB}=4\lambda^2_0\lambda^2_2$, and $\tau_{AC}=4\lambda^2_0\lambda^2_3$.
According to the monogamy relation (\ref{m4}), we have
\begin{eqnarray}\label{mu}
\mu_\tau= 1+\left(\frac{\lambda_4}{\lambda_3}\right)^2,
\end{eqnarray}
with $\lambda_2\geq \lambda_3\geq \lambda_4\geq0$. Thus, we obtain $\mu_\tau=\min\{1+\left(\lambda_4/\lambda_3\right)^2\}=1$, where the minimum is taken over all the states in (\ref{ex}). It can be verified that the W-type states ($\lambda_4=0$ in (\ref{ex})) saturate  the inequality.
For another quantum entanglement measure, we consider the concurrence defined by $C(|\psi\rangle_{AB})=\sqrt{\tau(|\psi\rangle_{AB})}=\sqrt{{2\left[1-\mathrm{Tr}(\rho_A^2)\right]}}$ for a bipartite pure state $|\psi\rangle_{AB}$. The concurrence of a mixed state is given by the convex roof extension $C(\rho_{AB})=\min_{\{p_i,|\psi_i\rangle\}}\sum_ip_iC(|\psi_i\rangle)$. Consider the state (\ref{ex}). One obtains
\begin{eqnarray}\label{mu1}
\mu_C= \sqrt{1+\left(\frac{\lambda_2}{\lambda_3}\right)^2+\left(\frac{\lambda_4}{\lambda_3}\right)^2}-\frac{\lambda_2}{\lambda_3}.
\end{eqnarray}
Let $f(x,y)=\sqrt{1+x^2+y^2}-x$, with $x=\frac{\lambda_2}{\lambda_3}\geq 1$ and $y=\frac{\lambda_4}{\lambda_3}\in [0,1]$. We obtain that $\mu_C=f_{\min}(x,y)\leq f_{\min}(1,0)= \sqrt{2}-1$. Therefore, in this case, $C\preceq \tau$ is $\mu_C< \mu_\tau$.

In \cite{ZXN}, the authors proved that the $\alpha$ power of concurrence for the $W$-class states does not satisfy the monogamy inequality (\ref{m1}) for $1\leq\alpha<2$. From the monogamy relation (\ref{m4}), we can obtain $C^\alpha_{A|BC}\geq (\sqrt{2}-1)^\alpha C^\alpha_{AB}+C^\alpha_{AC}$ for $W$-class states for $1\leq\alpha<2$, which solves the problem raised in Ref. \cite{ZXN}.
Moreover, for states beyond qubits, the monogamy inequality (\ref{m1}) does not apply for concurrence. For example, for the three-qutrit state \cite{ooo}, $|\Psi\rangle_{ABC}=\frac{1}{\sqrt{6}}(|123\rangle-|132\rangle+|231\rangle-|213\rangle+|312\rangle-|321\rangle)$, one has $C_{AB}=C_{AC}=1$ and $C_{A|BC}=\frac{4}{3}$, which violates (\ref{m1}). Nevertheless, from (\ref{m4}) and (\ref{theorem1}), we find that $|\Psi\rangle_{ABC}$ satisfies the monogamy relations for $\mu_C=\frac{1}{3}$ or $\alpha=5$.

The monogamy property depends on both the entanglement measure and quantum states. Some special classes of states, for example, the generalized $n$-partite GHZ-class states admitting the multipartite Schmidt decomposition \cite{gjl,ghz}, $|\psi\rangle_{A_1A_2\cdots A_n}=\sum_{i=1}\lambda_i|i_1\rangle\otimes|i_2\rangle\otimes\cdots \otimes|i_n\rangle$, $\sum_i\lambda_i^2=1$, $\lambda_i>0$, always satisfy monogamous relations for any entanglement measures, since one always has $E(|\psi\rangle_{A_1|A_2\cdots A_n})>0$ and $E(\rho_{A_1A_i})=0$ for all $i=2,\cdots,n$ and any entanglement measure $E$. For a general entanglement measure $E$, such as concurrence, the CKW inequality (\ref{m1}),
is usually not satisfied, whereas relation (\ref{m4}) holds. We show the connection between the CKW inequality (\ref{m1}) and relation (\ref{m4}) in Fig. 2.
For convenience, we denote by $\rho^{\otimes m}_{A_1A_2...A_n}=\rho_{A_{11}A_{12}...A_{1m}A_{21}...A_{2m}...A_{n1}...A_{nm}}$ and $\rho^{\otimes m}_{A_1A_i}=\rho_{A_{11}...A_{1m}A_{i1}...A_{im}}$, where $A_{ij}$ is the $i$th party of the $j$th copy of $\rho_{A_1A_2...A_n}$.
Let $E(\rho^{\otimes m}_{A_1|A_2...A_n})=E(\rho_{A_{11}A_{12}...A_{1m}|A_{21}...A_{2m}...A_{n1}...A_{nm}})$ denote the entanglement of the $m$ copies of $\rho_{A_1A_2...A_n}$ between the first party and the other ones, i.e., the entanglement between $A_{11}A_{12}...A_{1m}$ and $A_{21}...A_{2m}...A_{n1}...A_{nm}$; then $E(\rho^{\otimes m}_{A_1A_i})=E(\rho_{A_{11}...A_{1m}|A_{i1}...A_{im}})$ is the entanglement between the first party and the $i$th party.
We consider entanglement measures that are nonadditive because for an additive entanglement measure $E$, one trivially obtains $E(\rho\otimes \delta)=E(\rho)+E(\delta)$.

\begin{figure}
\centering
    \includegraphics[width=8.5cm]{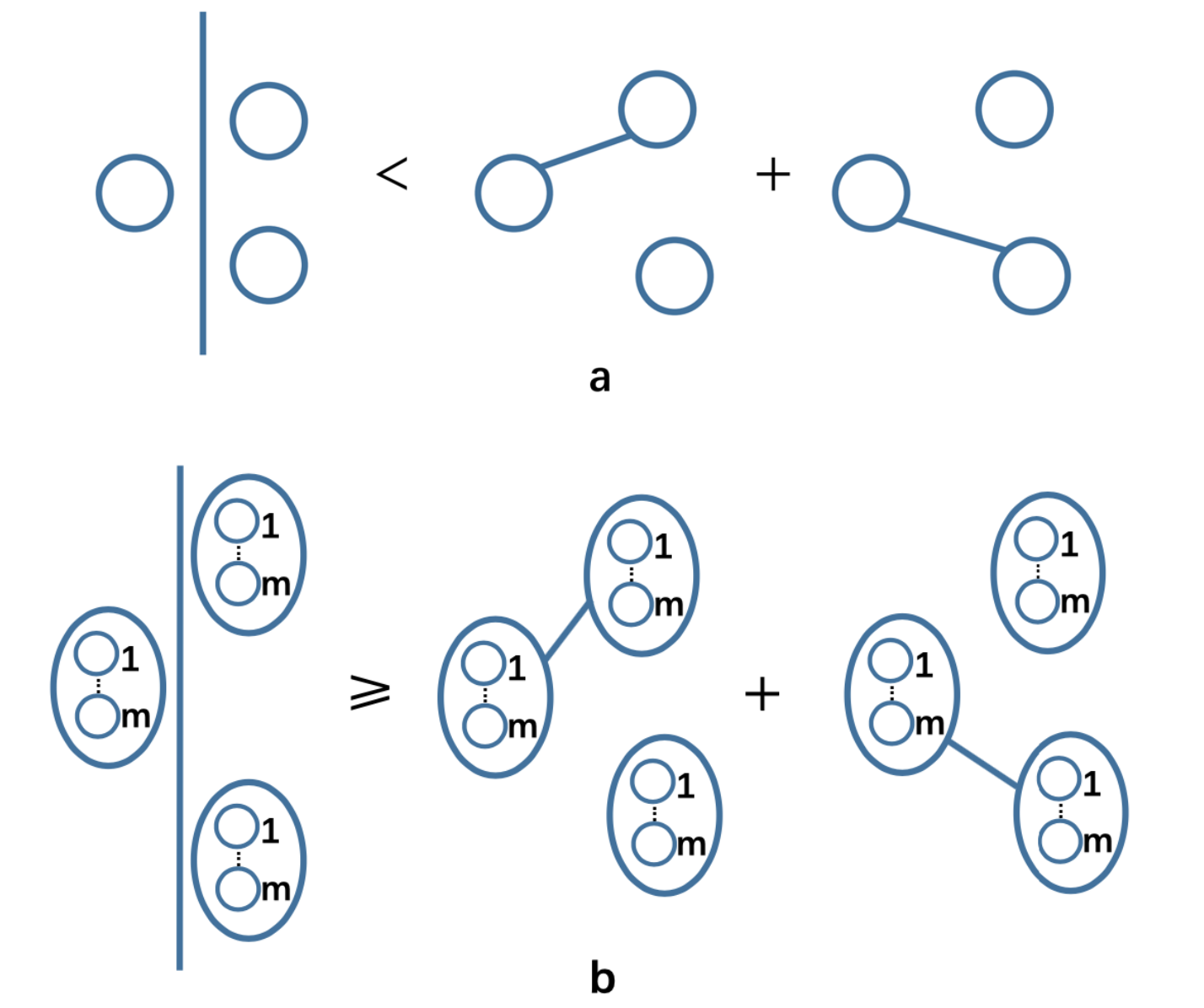}\\
\caption{a) The entanglement $E$ with respect to a tripartite state $\rho_{ABC}$ violates the CKW inequality (\ref{m1}), which corresponds to the regions with $0<\mu<1$ in Fig. 1.
b) Certain $m$-copy states of $\rho_{ABC}$ satisfy the monogamy relation; namely, the states corresponding to the orange, yellow, and white regions in Fig. 1 may move to the blue region under $m$ copies.}\label{2}
\end{figure}

{\bf Theorem 2}  A nonadditive measure of entanglement $E$ is monogamous according to definition (\ref{m4}) if and only if there exists an integer $m (m\geq1)$ such that
\begin{eqnarray}\label{th2}
E(\rho^{\otimes m}_{A|BC})\geq E(\rho^{\otimes m}_{AB})+E(\rho^{\otimes m}_{AC})
\end{eqnarray}
for any state $\rho_{ABC}$.

See Appendix for the proof of Theorem 2.

{\bf Corollary 1} For a nonadditive entanglement measure $E$ satisfying (\ref{m4}), but not the monogamy relation (\ref{m1}), there exists an integer $m$ $(m\geq1)$ such that
\begin{eqnarray}\label{coro}
E(\rho^{\otimes m}_{A|BC})\geq E(\rho^{\otimes m}_{AB})+E(\rho^{\otimes m}_{AC})
\end{eqnarray}
for any tripartite state $\rho_{ABC}$.

From definition (\ref{m4}) and the derivation of Theorem 2, we have $\mu=\frac{1-E_{AC}/E_{A|BC}}{E_{AB}/E_{A|BC}}$. Either $E_{AC}/E_{A|BC}$ or $E_{AB}/E_{A|BC}$ should increase when $\mu$ decreases. In other words, $E_{AC}/E_{A|BC}$ or $E_{AB}/E_{A|BC}$ increases as $\mu$ decreases. Therefore, the minimal number of required copies increases to reactivate the monogamy relationship. In other words, the monogamy weight $\mu$ and the integer $m$ are roughly inversely proportional.
For entanglement measures $E$, such as the tangle $\tau$ \cite{ckw}, whose monogamy weight is $\mu_\tau\geq 1$, that is, $\tau_{A|BC}\geq \kappa\tau_{AC}+\tau_{AB}~(0<\kappa\leq \mu_\tau)$, for any three qubit states, it is obvious that $m_\tau=1$.
Concerning measures $E$ such that $0<\mu<1$, let us consider concurrence $C$. Since the negativity
$N(|\psi\rangle_{AB})=\||\psi\rangle\langle\psi|^{T_B}\|-1=2\sqrt{\lambda_0\lambda_1}
=\sqrt{2(1-\mathrm{Tr}\rho_A^2)}=C(|\psi\rangle_{AB})$, where $|\psi\rangle_{AB}=\sqrt{\lambda_0}|00\rangle+\sqrt{\lambda_1}|11\rangle$, by using the results in \cite{jinzx,sen} we have $C(|W\rangle_{A|BC}^{\otimes m_C})=\frac{1}{2}[(1+\frac{4\sqrt{2}}{3})^{m_C}-1]$, $C(\rho_{AB}^{\otimes m_C})=C(\rho_{AC}^{\otimes m_C})=\frac{1}{2}[(1+\frac{4}{3})^{m_C}-1]$ for $m$ copies of the $W$ states, $|W\rangle_{ABC}=\frac{1}{\sqrt{3}}(|100\rangle+|010\rangle+|001\rangle)$, where $\rho_{AB}$ and $\rho_{AC}$ are the reduced states of $|W\rangle_{ABC}$. It can be observed that $C(|W\rangle_{A|BC}^{\otimes m_C})\geq C(\rho_{AB}^{\otimes m_C})+C(\rho_{AC}^{\otimes m_C})$ if $m_C\geq 4$. Here, the monogamy weight of $|W\rangle_{ABC}$ for the concurrence is $\mu_C=\sqrt{2}-1$ from (\ref{mu1}). Therefore, in this case, we have $C\preceq \tau$ as $\mu_C< \mu_\tau$, whereas $m_C\geq m_\tau$.

The monogamy relation defined in (\ref{m4}) can be generalized to multipartite systems. For any $N$-partite state $\rho_{AB_1\cdots B_{N-1}}$, we obtain the following result if $E$ satisfies (\ref{m4}) for any tripartite state (see the proof in the appendix).

{\bf Theorem 3 }.  Assume that, for any $N$-partite state $\rho_{AB_1\cdots B_{N-1}}$, ${E_{AB_i}}\geq {E_{A|B_{i+1}\cdots B_{N-1}}}$ for $i=1, 2, \cdots, m$, and
${E_{AB_j}}\leq {E_{A|B_{j+1}\cdots B_{N-1}}}$ for $j=m+1,\cdots,N-2$, $\forall$ $1\leq m\leq N-3$, $N\geq 4$. If $E$ satisfies relation (\ref{m4}), then for the tripartite states, we have
\begin{eqnarray*}\label{th1}
&&E_{A|B_1B_2\cdots B_{N-1}}\nonumber \\
&&\geq E_{AB_1}+\Gamma_1E_{AB_2}+\cdots+\Gamma_{m-1}E_{AB_m}\nonumber\\
&&+\Gamma_m(\mu_{m+1}E_{AB_{m+1}}+\cdots+\mu_{N-2}E_{AB_{N-2}}+E_{AB_{N-1}}),
\end{eqnarray*}
where $\Gamma_k=\Pi_{i=1}^k\mu_i$, $k=1,2,\cdots,N-2$, and $\mu_i$ denotes the monogamy weight of the $(N+1-i)$-partite state $\rho_{AB_1\cdots B_{N-i}}$.

In Theorem 3 we have assumed that some ${E_{AB_i}}\geq {E_{A|B_{i+1}\cdots B_{N-1}}}$ and some
${E_{AB_j}}\leq {E_{A|B_{j+1}\cdots B_{N-1}}}$ for the $N$-partite state $\rho_{AB_1\cdots N_{N-1}}$.
If all ${E_{AB_i}}\geq {E_{A|B_{i+1}\cdots B_{N-1}}}$ for $i=1, 2, \cdots, N-2$, then we have $E_{A|B_1\cdots B_{N-1}}=E_{AB_1}+\Gamma_1E_{AB_2}+\cdots+\Gamma_{N-2}E_{AB_{N-1}}$.

The monogamy weight $\mu$ functions as a bridge for characterizing the monogamous ability of different entanglement measures. An entanglement measure $E$ is more likely to become monogamous as $\mu$ increases. Then, $\mu$ provides the physical meaning of the coefficients introduced in Ref. \cite{jzx} for the weighted monogamy relations.
Furthermore, monogamy has emerged as an ingredient in the security analysis of quantum key distributions \cite{mp}. From Theorem 2, we can see that the monogamy relation can be reactivated by finite $m$ copies of $\rho$ for nonadditive entanglement measures. In other words, they can still be used for secure communication against individual attacks by the eavesdropper by reactivating the monogamy property of $\rho$.

Thus, entanglement monogamy is a fundamental property of multipartite systems. We introduced a new definition of the relation for entanglement measures that characterizes the precise division of the entanglement distribution for a given entanglement measure, $E$ (Fig. 1). The non-monogamous entanglement distribution is only located on the boundary of the square (except for the coordinate axis): the blue region for both our notion of monogamy (\ref{m4}) and the conventional one (\ref{m1}), whereas the orange, yellow, and white regions violate (\ref{m1}), but still work for our notion of monogamy (\ref{m4}). Our definition of monogamy is based on equality (\ref{m4}) rather than previous inequalities (\ref{m1}). The advantage of our notion of monogamy is that one can distinguish which entanglement measure is more easily monogamous by comparing the monogamy weights. We have used concurrence and tangle as examples, showing that tangle is more likely monogamous than concurrence because the weight of the tangle is larger than that of concurrence, corresponding to previous results \cite{ckw,ak}. However, using $E^\alpha$ for some $\alpha>0$, we have shown that the definition of our monogamy relation can reproduce conventional monogamy inequalities such as (\ref{m1}). We then showed that every nonadditive entanglement measure that violates the conventional monogamy inequalities but satisfies our definition can be recovered as monogamous if one allows for many copies of the state, that is, corresponding to the orange, yellow, and white regions in Fig. 1. Our definition can also be generalized to multipartite systems. Theorem 3 provides a general relation for $N$-partite states. Our results may shed light on monogamy properties related to other quantum correlations.

\bigskip
\noindent{\bf Acknowledgments}\, \,
This work was supported in part by the National Natural Science Foundation of China (NSFC) under Grants 11847209, 12075159, 11975236, 12075159, and 11635009; Beijing Natural Science Foundation (Grant No. Z190005), Academy for Multidisciplinary Studies, Capital Normal University, Shenzhen Institute for Quantum Science and Engineering, Southern University of Science and Technology (No. SIQSE202001), Academician Innovation Platform of Hainan Province, China Postdoctoral Science Foundation funded project No. 2019M650811, and China Scholarship Council No. 201904910005.

\bigskip
\section*{APPENDIX}
\setcounter{equation}{0} \renewcommand%
\theequation{A\arabic{equation}}

\subsection{Proof of Theorem 1}

Let $E$ be a monogamous measure of the entanglement that satisfies (\ref{m4}). If $E_{A|BC}=0$, the result is clear. We assume $E_{A|BC}> 0$. As $E$ is a measure of quantum entanglement, it is non-increasing under a partial trace, and $E_{AC}\geq E_{AB}$ for $\mu>0$ according to (\ref{m4}). We have $E_{A|BC}> E_{AC}\geq E_{AB}$ for any state $\rho_{ABC}$. Set $x_1=E_{AB}/E_{A|BC}\in [0,1)$ and $x_2=E_{AC}/E_{A|BC}\in [0,1)$. Clearly, $\gamma$ exists such that
\begin{eqnarray}\label{pfth11}
x_1^\gamma+x_2^\gamma\leq 1,
\end{eqnarray}
since $x_1^\gamma$ and $x_2^\gamma$ decrease when $\gamma$ increases.
Set $f(\rho_{ABC}):=\inf_{\gamma}\{\gamma|x_1^\gamma+x_2^\gamma\leq 1, x_1,x_2\in [0,1)\}$.
Owing to the compactness of the set of tripartite states $\rho_{ABC}$ and the continuity of $E$,
there always exists a state $\rho_1$, such that $\max_{\rho} x_1(\rho)=x_1(\rho_1)<1$ for $x_1(\rho)\in [0,1)$. Similarly, there exists a state $\rho_2$ such that $\max_{\rho} x_2(\rho)=x_2(\rho_2)<1$. Therefore, there exists a sufficiently large $N$ independent of $\rho_{ABC}$, such that $x_1^N(\rho)+x_2^N(\rho)\leq 1$ as $x_i^N(\rho)\to 0$, $i=1,2$, for a sufficiently large $N$. Thus, we can always obtain a sufficiently large $N$ independent of $\rho$, such that $x_1^N(\rho_0)+x_2^N(\rho_0)\leq x_1^N(\rho_1)+x_2^N(\rho_2)\leq 1$, where $x_i(\rho_i)=\max_{\rho} x_i(\rho)<1$, $i=1,2$.

Next, we prove that $f(\rho_{ABC})$ is bounded uniformly. It is only necessary to prove that $f(\rho_{ABC})\leq N$ for any $\rho_{ABC}$. If there exists a state $\rho_0$ such that $f(\rho_0)>N$, then on one hand, by the definition of $f(\rho_{ABC})$, i.e., $x_1^{f(\rho_0)}(\rho_0)+x_2^{f(\rho_0)}(\rho_0)\leq1$, one has $x_1^N(\rho_0)+x_2^N(\rho_0)>1$ as $f(\rho_0)>N$. In contrast, $x_1^N(\rho_0)+x_2^N(\rho_0)\leq x_1^N(\rho_1)+x_2^N(\rho_2)\leq 1$, where $x_i(\rho_i)=\max_{\rho} x_i(\rho)<1,~i=1,2$, which gives rise to a contradiction. Therefore, $f(\rho_{ABC})$ is bounded uniformly. Setting $\alpha=\sup_{\rho_{ABC}}f(\rho_{ABC})$, we prove the inequality (\ref{theorem1}).

Moreover, without loss of generality, we assume that $E_{AC}\geq E_{AB}$ from Eq. (\ref{theorem1}), if $E_{A|BC}=E_{AC}$, then $E_{AB}=0$ for the pure state $|\psi\rangle_{ABC}$. Otherwise, $E_{A|BC}> \max\{E_{AB}, E_{AC}\}$. Obviously, in any case, there always exists a constant $\mu$ such that $E_{A|BC}= \mu E_{AB}+E_{AC}$ because we can always choose $\mu=\min_{\rho_{ABC}}\frac{E_{A|BC}- E_{AC}}{E_{AB}}$ for the latter case.

\subsection{Proof of Theorem 2}

Let $\lambda_i$ and $|i\rangle$ be the eigenvalues and eigenstates of state $\rho_{AB'}$ in systems $A$ and $B'$, respectively. We can always introduce a third system $B''$. The systems $B'$ and $B''$ together constitute the system $B$. Provided the dimension of the system $B''$ is not smaller than that of $AB'$, there exists an orthonormal basis $|\hat{i}\rangle$ of $B''$ such that $|\psi\rangle=\sum_i\sqrt{\lambda_i}|i\rangle|\hat{i}\rangle$ is a pure state of the tripartite system $AB'B''$.
Thus $\rho_{AB'}=\mathrm{tr}_{B''}|\psi\rangle\langle\psi|$, where $\mathrm{tr}_{B''}$ is the partial trace over $B''$. As $\mathrm{tr}_{B''}$ is a local operation performed on $B''$, one has $E(\rho_{AB'})\leq E(|\psi\rangle\langle\psi|)$. As $\rho_A=\mathrm{tr}_{B}|\psi\rangle\langle\psi|=\mathrm{tr}_{B'}\rho_{AB'}$, the Schmidt coefficients of $|\psi\rangle$ are $\sqrt{\lambda_i(\rho_A)}$. Hence, the quantum entanglement has the form
\begin{eqnarray*}
E(|\psi\rangle\langle\psi|)=f(\vec{\lambda}(\rho_A)),
\end{eqnarray*}
where $f$ is a function of $\vec{\lambda}(\rho_A)$ given by the nonzero eigenvalues of state $\rho_A$. Thus, $E(\rho_{AB'})$ depends only on $\rho_A$ and $E(\rho_{AB'})\leq f(\vec{\lambda}(\rho_A))$. Therefore, there exists a positive number $0\leq L\leq 1$, such that $E(\rho_{AB'})=L f(\vec{\lambda}(\rho_A))$.
Note that $\rho_{AB'}^{\otimes m}$ is shortest for $\rho_{A_1B'_1}\otimes \rho_{A_ 2 B '_2}\otimes \cdots\otimes \rho_{A_mB'_m}$. The eigenvalues of $\rho_{AB'}^{\otimes m}$ are $\{\Pi_{j=1}^m \lambda_{i_j}\}$. Hence, a function $g$ exists such that $E(\rho_{AB'}^{\otimes m})=g^m[E(\rho_{AB'})]$. Then, $E(\rho_{AB'}^{\otimes m})=g^m[L f(\vec{\lambda}(\rho_A))]=L^mg^m[f(\vec{\lambda}(\rho_A))]$.
Similar to $E(\rho_{AB'})$, assuming that $E(\rho_{AB''})=M f(\vec{\lambda}(\rho_A))$ with $0\leq M\leq 1$, we have $E(\rho_{AB''}^{\otimes m})=M^m g^m[f(\vec{\lambda}(\rho_A))]$ and $E(\rho^{\otimes m}_{A|B'B''})=g^m[f(\vec{\lambda}(\rho_A))]$.
According to definition (4), one gets that $M=0$ if $L=1$ and $L=0$ if $M=1$ for the pure state $|\psi\rangle$. Otherwise, $E(\rho^{\otimes m}_{A|B'B''})> \max\{E(\rho^{\otimes m}_{AB'}), E(\rho^{\otimes m}_{AB''})\}$.

We consider the cases of $0<L<1$ and $0<M<1$. To obtain (\ref{th2}), it is sufficient to determine the minimum integer $m$ such that
\begin{eqnarray}\label{pfth31}
{L}^m+{M}^m\leq1.
\end{eqnarray}
From Theorem 1, there always exists a positive integer $m$ such that inequality (\ref{pfth31}) holds.

In the following, we prove that if $E$ is monogamous in tripartite pure states $|\psi\rangle$, then it is also monogamous in tripartite mixed states.
Let $\rho_{ABC}=\sum_ip_i|\psi_i\rangle\langle \psi_i|_{ABC}$ be the optimal decomposition such that $E(\rho_{ABC})=\sum_ip_iE(|\psi_i\rangle_{ABC})$. Denote ${\rho_i}_{AB}=\mathrm{Tr}_C(|\psi_i\rangle\langle \psi_i|_{ABC})$ and ${\rho_i}_{AC}=\mathrm{Tr}_B(|\psi_i\rangle\langle \psi_i|_{ABC})$.
We have
\begin{eqnarray*}
E(\rho_{ABC})&&=\sum_ip_iE(|\psi_i\rangle_{ABC})\\
&&\geq\sum_ip_iE({\rho_i}_{AB})+\sum_ip_iE({\rho_i}_{AC})\\
&&\geq E({\rho}_{AB})+E({\rho}_{AC}),
\end{eqnarray*}
where the last inequality follows from the convexity of measure $E$, ${\rho}_{AB}=\sum_ip_i{\rho_i}_{AB}$, and ${\rho}_{AC}=\sum_ip_i{\rho_i}_{AC}$.

\bigskip

\subsection{Proof of Theorem 3}

From (\ref{m4}), we have
\begin{eqnarray}\label{pf41}
&&E_{A|B_1B_2\cdots B_{N-1}}\nonumber\\
&&= E_{AB_1}+\mu_1E_{A|B_2\cdots B_{N-1}}\nonumber\\
&&= E_{AB_1}+\mu_1 E_{AB_2}+\mu_1\mu_2E_{A|B_3\cdots B_{N-1}}\nonumber\\
&&= \cdots\nonumber\\
&&= E_{AB_1}+\mu_1 E_{AB_2}+\cdots+\mu_1\mu_2\cdots \mu_{m-1}E_{AB_m}\nonumber\\
&&~~~+\mu_1\mu_2\cdots \mu_mE_{A|B_{m+1}\cdots B_{N-1}}.
\end{eqnarray}
Similarly, because ${E_{AB_j}}\leq {E_{A|B_{j+1}\cdots B_{N-1}}}$ for $j=m+1,\cdots,N-2$, we obtain
\begin{eqnarray}\label{pf42}
&&E_{A|B_{m+1}\cdots B_{N-1}} \nonumber\\
&&= \mu_{m+1}E_{AB_{m+1}}+E_{A|B_{m+2}\cdots B_{N-1}}\nonumber\\
&&= \mu_{m+1}E_{AB_{m+1}}+\cdots+\mu_{N-2}E_{AB_{N-2}}\nonumber\\
&&~~~+E_{AB_{N-1}}.
\end{eqnarray}
By combining (\ref{pf41}) and (\ref{pf42}), we have Theorem 3.

\end{document}